\documentclass[fleqn,usenatbib]{mnras}

\usepackage{newtxtext,newtxmath}

\usepackage[T1]{fontenc}
\usepackage{dcolumn}
\usepackage{bm}

\usepackage{etoolbox}
\usepackage{threeparttable}
\usepackage{array}
\usepackage{xspace}
\usepackage{multirow}
\usepackage{booktabs}
\usepackage[version=4]{mhchem}
\usepackage{xcolor}
\usepackage{natbib}
\usepackage{hyperref}

\newcolumntype{d}{D{.}{.}{5}}
\newcolumntype{e}{D{.}{.}{10}}
\newcommand{\wn}{cm$^{-1}$\xspace} 
\newcommand{\mrm}[1]{\ensuremath{\mathrm{#1}}}

\DeclareRobustCommand{\VAN}[3]{#2}
\let\VANthebibliography\thebibliography
\def\thebibliography{\DeclareRobustCommand{\VAN}[3]{##3}\VANthebibliography}

\usepackage{graphicx}	
\usepackage{amsmath}	

\title[Collisional excitation of \ce{PO+} by \emph{para}-\ce{H2}]{Collisional excitation of \ce{PO+} by \emph{para}-\ce{H2}: Potential Energy Surface, Scattering Calculations and Astrophysical Applications}

\author[F. Tonolo et al.]{
F. Tonolo,$^{1,2}$\thanks{E-mail: francesca.tonolo@sns.it}
L. Bizzocchi,$^{2}$\thanks{E-mail: luca.bizzocchi@unibo.it}
V. M. Rivilla, $^{3}$
F. Lique, $^{4}$
M. Melosso $^{2}$
and C. Puzzarini$^{2}$
\\
$^{1}$Scuola Normale Superiore, Piazza dei Cavalieri 7, I-56126 Pisa, Italy.\\
$^{2}$Dipartimento di Chimica “Giacomo Ciamician”, Università di Bologna, Via F. Selmi 2, I-40126 Bologna, Italy.\\
$^{3}$Centro de Astrobiología (CAB), INTA-CSIC, Carretera de Ajalvir km 4, Torrejón de Ardoz, 28850, Madrid, Spain.\\
$^{4}$Univ. Rennes, CNRS, IPR (Institut de Physique de Rennes) – UMR 6251, F-35000 Rennes, France.\\
}

\date{Accepted XXX. Received YYY; in original form ZZZ}

\pubyear{2023}

\begin{document}
\label{firstpage}
\pagerange{\pageref{firstpage}--\pageref{lastpage}}
\maketitle

\begin{abstract}
We report the derivation of rate coefficients for the rotational (de-)excitation of \ce{PO+} induced by collisions with \ce{H2}.
The calculations were performed on a four-dimensional potential energy surface, obtained on top of 
highly accurate \emph{ab initio} energy points. Preliminary tests pointed out the low influence of the coupling between $j=0$ and the higher rotational levels of \ce{H2} on the cross sections values, thus allowing to neglect the rotational structure of \ce{H2}.
On this basis, state-to-state collisional rate coefficients were derived for temperatures ranging from 5 to 200\,K.
Radiative transfer calculations have been used to model the recent observation of  \ce{PO+} in the G+0.693-0.027 molecular cloud, in order to evaluate the possible impact of non-LTE models on the determination of its physical conditions.
The derived column density was found to be approximately $\sim$\,$3.7\times10^{11}$ cm$^{-2}$, which is 60\% (a factor of $\sim$\,$1.7$) smaller than the previously LTE-derived value.
Extensive simulations show that \ce{PO+} low-$j$ rotational lines exhibit maser behavior at densities between $10^4$ and $10^6$\,cm$^{-3}$, thus highlighting the importance of a proper treatment of the molecular collisions to accurately model \ce{PO+} emissions in the interstellar medium.

\end{abstract}




\section{Introduction}
\indent\indent
The investigation of the cosmic abundance and distribution of phosphorus (\ce{P}) in space deserves a special attention as it is considered a biogenic element together with carbon, hydrogen, oxygen, nitrogen and sulfur (CHONPS, \citealt{bergner2022astrochemistry,oberg2021astrochemistry, rivilla2016first}).
In particular, \ce{P} has a pivotal importance for ``abiogenesis'', namely, the formation of prebiotic species from abiotic systems \citep{pearce2017origin}. It is indeed a key ingredient for the composition of many biomolecules, especially when bonded with \ce{O} atoms in the form of phosphate (\ce{PO4^{3-}}). 
For this reason, \ce{P} is an ubiquitous element in our planet and its abundance in living organisms is relatively high \citep{fagerbakke1996content}.
Outside Earth, P-bearing compounds have been found in a variety of environments, from the planetary atmospheres of Jupiter and Saturn \citep{bregman1975observation,ridgway1976800} to meteorites \citep{pasek2005aqueous,schwartz2006phosphorus}, as well as in the 67P/Churyumov-Gerasimenko comet \citep{altwegg2016prebiotic,rivilla2020alma} and in circumstellar envelopes of evolved stars \citep{agundez2014confirmation,halfen2008detection,agundez2007discovery,tenenbaum2007identification, rivilla2020alma}. 
Recently, P has also been found in Enceladus’s ocean in the form of orthophosphates, opening a new window on the origin of life under the frozen surfaces of Jupiter's moons \citep{postberg2023detection}.

All this contrasts with its actual, limited distribution in the interstellar medium (ISM), where only a few P-bearing molecules have been identified (see \citealt{rivilla2022ionize} and references therein), despite numerous searches (e.g. \citealt{chantzos2020first}). 
Such elusiveness may be due to the high sublimation temperature of atomic \ce{P}, which leads to a strong depletion of this element onto interstellar grains \citep{lebouteiller2005phosphorus}. 
The first P-bearing species detected in the ISM were \ce{PN} \citep{ziurys1987detection,turner1987detection} and the \ce{CP} radical \citep{guelin1990free}, whereas the possible precursor of this latter, \ce{HCP}, was observed almost 20 years later \citep{agundez2007discovery}. 
From 2007 to date, few other molecules have been observed, namely \ce{PO} \citep{lefloch2016phosphorus,rivilla2016first, rivilla2018phosphorus, bergner2019detection}, \ce{C2P} \citep{halfen2008detection}, \ce{PH3} \citep{agundez2014confirmation} and, very recently, \ce{PO+} \citep{rivilla2022ionize}.

Besides the paucity of observational data, only few studies on \ce{P}-reactivity \citep{de2021formation,viana2009quantum,baptista2023phosphine,alessandrini2021search} and chemical modelling \citep{fontani2016phosphorus,rivilla2016first, lefloch2016phosphorus,jimenez2018chemistry,chantzos2020first,rivilla2022ionize} are present in the literature, thus making the understanding of the \ce{P} chemistry in the ISM far from being satisfactory.
Overcoming this lack of information requires a major astrochemical effort, and the first step is achieving new detections of P-bearing species complemented by a reliable determination of their abundances. 
This latter aspect calls for extra caution for environments where local thermodynamic equilibrium (LTE) conditions may not be fulfilled, and radiative transfer calculations should be undertaken.
Under such conditions, the estimate of molecular abundances from spectral lines requires the knowledge of the collisional rate coefficients of the target species with the most abundant perturbing gas, i.e., molecular hydrogen (\ce{H2}), with \ce{He} sometimes being considered as approximation for \textit{para}-\ce{H2} ($j = 0$) \citep{roueff2013molecular}.

An interesting case is given by the recent detection of \ce{PO+} in the molecular cloud G+0.693-0.027 \citep{rivilla2022ionize}, located in the SgrB2 region of the center of the Galaxy, where other P-bearing species where previously detected \citep{rivilla2018phosphorus}. This source is characterized by a \ce{H2} gas density of several $1\times10^{4}$\,cm$^{-3}$ \citep{zeng2020cloud}. Due to this relatively low density, the LTE conditions are not achieved, and hence the energy levels of the molecules are not thermalized at the kinetic temperature of the cloud ($\sim$\,150\,K; \citet{zeng2018complex}). This motivates the need of collisional rate coefficients to properly describe the molecular excitation of \ce{PO+}. Recently,  
\citet{chahal2023po+} investigated the collisional behavior of \ce{PO+} with \ce{He} and provided the first set of collisional coefficients for non-LTE modelling of the abundance of \ce{PO+} in the ISM. 
However, for molecular hydrides and ions, \ce{He} does not represent a suitable template for collisions with \ce{H2} \citep{roueff2013molecular}. 

In order to provide collisional data that meet the astrophysical needs, we investigated the collision of \ce{PO+} with \textit{para}-\ce{H2}. 
These new data allowed us to test the reliability of the LTE approximation and to refine the column density value of \ce{PO+} obtained from the observations of the G+0.693-0.027 molecular cloud \citep{rivilla2022ionize}.  
This paper is organized as follows: \S~\ref{comp} provides the computational details 
--- for both the calculation of the interaction potential (\S~\ref{pes}) and dynamics (\S~\ref{scat}) --- to derive the collisional rate coefficients. \S~\ref{aa} assesses
the impact of the collisional data on the modelling of the column density of \ce{PO+} in G+0.693-0.027 cloud. 
Finally, in \S~\ref{concl}, the main outcomes of this investigation are presented.

\section{Computational details}\label{comp}
\begin{figure}
\centering
 \includegraphics[scale=0.32]{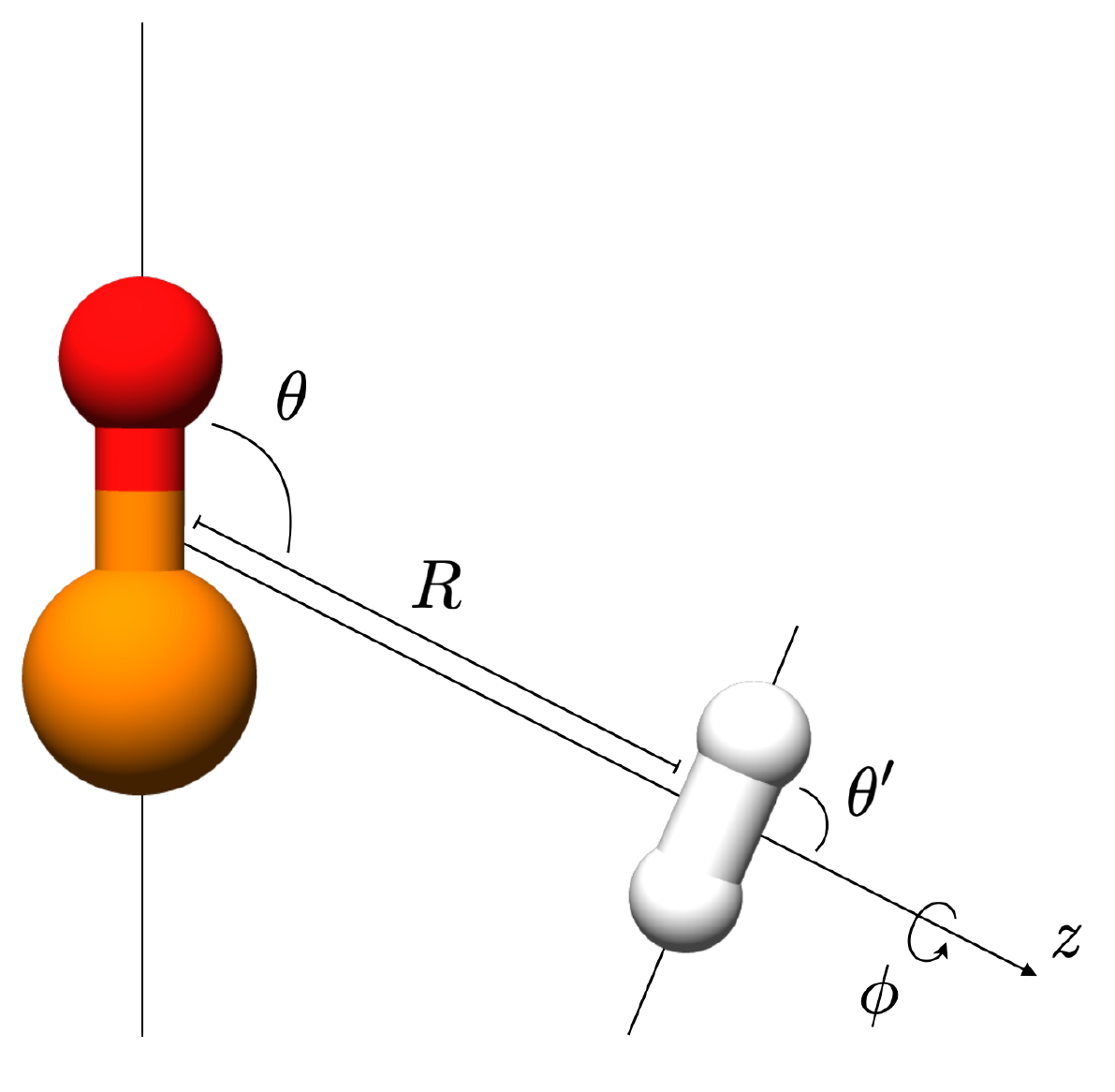}
 \caption{Jacobi internal coordinates of the \ce{PO+}$-$\,\ce{H2} collisional system. 
 }
 \label{fig1}
\end{figure}
\begin{figure*}  
 \includegraphics[scale=0.75]{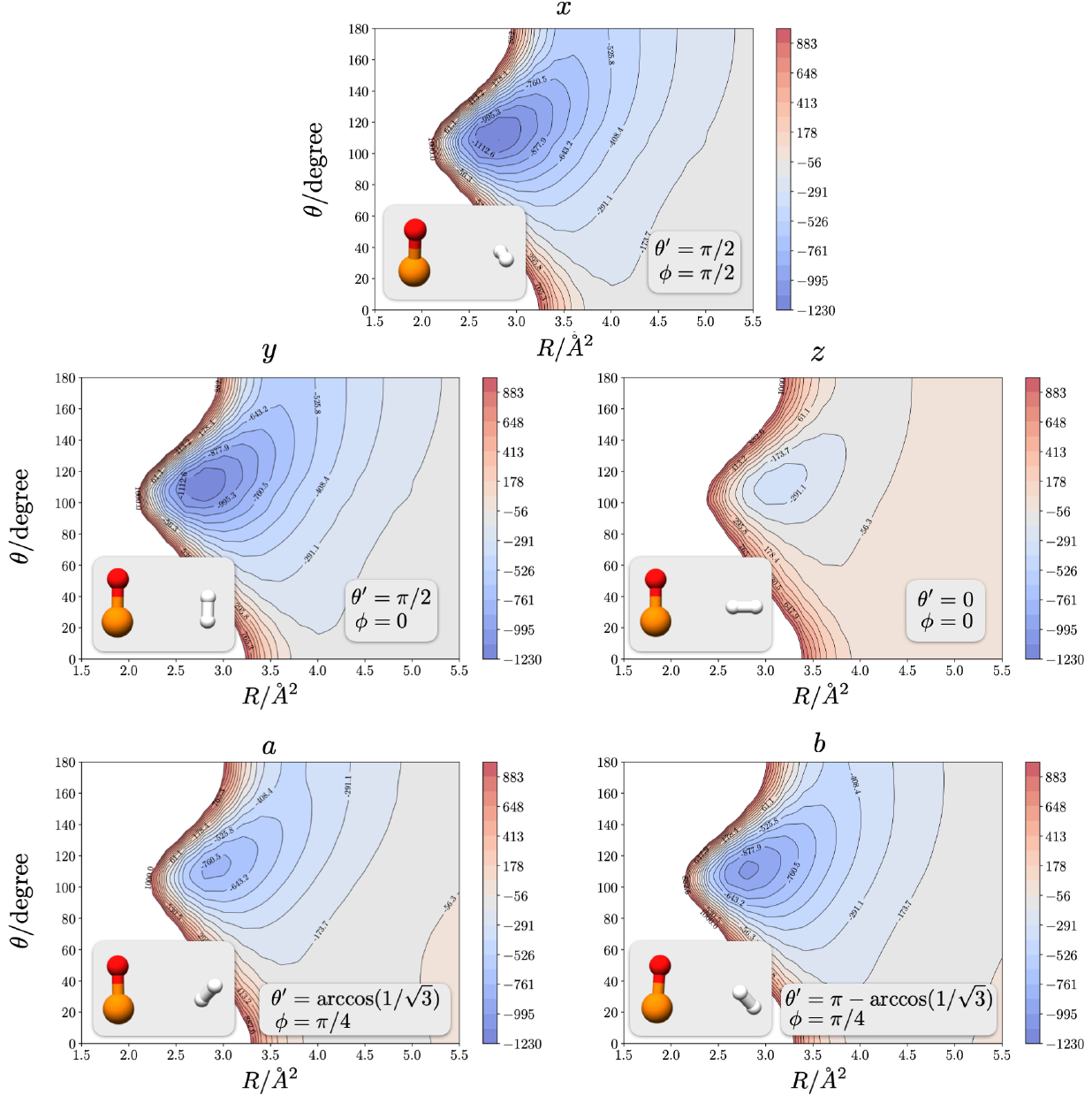}
 \caption{Contour plots of the \ce{PO+}--\ce{H2} interaction PES for five different orientations of \ce{H2}.}
 \label{fig2} 
\end{figure*}
The starting point to derive the collisional coefficients of a molecular system is the calculation of the interaction potential between the two colliding partners, in this case \ce{PO+} and \ce{H2}. 
This serves as a basis to solve the nuclear Schrödinger equation which describes the quantum scattering problem, thus providing the $\mathsf{S}$ matrix that contains all collisional information on the target system.
Both of these steps, detailed in the following two subsections, require extensive calculations that need to balance accuracy, suitability for physical applications, and computational efficiency.

\subsection{Potential energy surface}\label{pes}
\indent\indent
The interaction between \ce{PO+} and \ce{H2} has been described by a set of four Jacobi coordinates, as depicted in Figure\,\ref{fig1}.
These correspond to ($i$) the distance $R$ between the center of mass of \ce{PO+} and that of \ce{H2}, ($ii$) the angle $\theta$ between the molecular axis of \ce{PO+} and the vector $\textbf{R}$, and two angles, ($iii$) $\theta^{\prime}$ and ($iv$) $\phi$, defining the orientation of \ce{H2} in and out the plane formed by \ce{PO+} and vector $\textbf{R}$\@.

The interaction energies between the two collisional partners have been computed over a $\{R,\theta,\theta^{\prime},\phi\}$ grid, purposely chosen to accurately sample the anisotropy of the system. 
Moreover, \ce{PO+} and \ce{H2} were considered as rigid bodies, as we expect all the vibrational channels to be closed at the typical ISM physical conditions \citep{stoecklin2013ro}. 
The \ce{PO+} bond length was held fixed at its experimental equilibrium value \citep{petrmichl1991microwave}: $r(\ce{PO+})=1.4250$\,\AA. For \ce{H2}, we adopted the bond length corresponding to the averaged value over its ground vibrational state: $r_0(\ce{H2})=0.7667$\,\AA \,\citep{jankowski1998ab}. 

The electronic energy for each point of the $\{R,
\theta,\theta^{\prime},\phi\}$ grid has been computed 
using the explicitly correlated CCSD(T)-F12a method \citep{adler2007simple,knizia2009simplified,peterson2008systematically}, where the acronym stands for coupled cluster singles, doubles, and a perturbative treatment of triple excitations \citep{raghavachari1989fifth}, in conjuction  
with the aug-cc-pVQZ basis set augmented by an additional $d$ function on second-row atoms \citep{dunning2001gaussian, woon1993gaussian} (CCSD(T)-F12a/aug-cc-pV(Q+d)Z level of theory).  
The aug-cc-pV(Q+d)Z basis set has been chosen because the inclusion of diffuse functions (denoted by the aug- prefix) has proven to yield better performances in computing the electronic energies of charged systems, in which the electron density extends relatively far from the global maximum \citep{kendall1992electron,tonolo2021improved}.
For all the calculations, the MOLPRO suite of programs\footnote{\url{https://www.molpro.net}.} \citep{werner2012wires} has been employed.

The interaction energies were computed as the difference between the energy of the molecular complex ($E_\mrm{AB}$) and the sum of the energies of the two fragments $(E_\mrm{A}, E_\mrm{B})$.
All the terms have also been corrected for the basis set superposition error (BSSE) by means of the counterpoise (CP; \citealt{boys1970calculation}) correction scheme:
\begin{equation}
 \Delta E_{\text{CP}}= (E^{\mrm{AB}}_{\mrm{A}} - E^{\mrm{A}}_{\mrm{A}}) + (E^{\mrm{AB}}_{\mrm{B}} - E^{\mrm{B}}_{\mrm{B}})\,.
\end{equation}
Here, $E^{\mrm{AB}}_{\mrm{X}}$ is the energy of the monomer calculated with the same basis set used for the cluster and $E^{\mrm{X}}_{\mrm{X}}$ is the energy of the monomer computed with its own basis set ($\mrm{X} = \mrm{A},\mrm{B}$). 

To achieve an accurate characterization of the potential energy surface (PES) of the \ce{PO+}--\ce{H2} system, the coordinates of the \emph{ab initio} points were chosen in order to build up a dense mesh near the most anisotropic parts of the potential, whereas a coarser grid was adopted in regions where the energy mildly depends on the system geometry.
In order to further reduce the computational cost, we considered only five orientations of \ce{H2} with respect to \ce{PO+}, described by the $\{\theta^{\prime},\phi\}$ coordinates. 
This approximation has been found appropriate for similar systems (e.g.\  \ce{HCO+/H2}, see \citealt{tonolo2022hyperfine}) since the dependence of the potential on the orientation of \ce{H2} is very weak.

This statement deserves a more detailed note. In a two rigid rotor system, the interaction potential can be retrieved from \emph{ab initio} points by fitting them as an expansion over angular functions of the following form \citep{green1975rotational,wernli2007rotational,wernli2007rotationalb}:
\begin{equation}\label{fit}
 V\left(R,\theta,\theta',\phi\right) = \sum_{l_1 l_2 \mu} v_{l_1 l_2 \mu}(R) s_{l_1 l_2 \mu}\left(\theta,\theta',\phi\right) \,.
\end{equation}
Here, $v_{l_1 l_2 \mu}(R)$ are the radial coefficients and the $l_1$, $l_2$ and $\mu$ are indices associated with the rotational angular moments of \ce{PO+}$(j_1)$, \ce{H2}$(j_2)$ and their vector sum, respectively. $s_{l_1 l_2 \mu}$ are the angular coefficients, defined as products of spherical harmonics, $Y_{l_i m}(\theta^{\prime},\phi)$, and Clebsch-Gordan vector-coupling coefficients (the reader is referred to \citealt{green1975rotational,edmonds2016angular,brown2003rotational} for further details).
If we assume ${l_2 \leq 2}$, there are only four spherical harmonic functions that shape the dependence of the potential on each set of \{$\theta^{\prime},\phi$\}. Hence, the choice of five orientations of \ce{H2} with respect to \ce{PO+} not only suffices to describe the corresponding angular dependence of the potential, but also provides an over-determined system to test the accuracy of the ${l_2 \leq 2}$ truncation (see for details, \citealt{wernli2006collisions,tonolo2022hyperfine}).

The five orientations chosen ($x,y,z,a,b$) are the same as those used by \citet{wernli2006collisions} and \citet{tonolo2022hyperfine} for the \ce{HC3N\,/\,H2} and \ce{HCO+\,/\,H2} systems and are depicted in the insets of Figure~\ref{fig2}.
For each of them, 650~interaction energies were computed, spanning through 25 $\theta$ angle values equally spaced from 0 to 180 degrees and 26 $R$ distances, varying between 2\,\AA\, and 12\,\AA\,, and with a denser mesh between 2.6\,\AA\, and 3.6\,\AA.
Each set of $\{\theta^{\prime},\phi\}$ energies has been subsequently expressed as an expansion over $P_{\lambda}$ Legendre polynomials within the following expression \citep{lique2019gas}:
\begin{equation} \label{pot}
 V\left(R, \theta \right)=\sum_{\lambda}  v_{\lambda} (R) P_{\lambda} \left(\cos \theta \right) \,.
\end{equation}
The $v_{\lambda} (R)$ radial coefficients have been fitted to a functional form which takes into account the sizable contribution due to induction interactions of the \ce{PO+} ion:
\begin{multline} \label{vexp}
 v_\lambda(R) = \mrm{e}^{-a_1^\lambda R}\left(a_2^\lambda + a_3^\lambda R + a_4^\lambda R^2 + a_5^\lambda R^3\right) \\
 -\frac{1}{2}\left[1 + \tanh\left(R/R_\text{ref}\right)\right]
 \left(\frac{C^{\lambda}_4}{R^4} + \frac{C^{\lambda}_6}{R^6} + \frac{C^{\lambda}_8}{R^8} + \frac{C^{\lambda}_{10}}{R^{10}}\right) \,,
\end{multline}
where $a^{\lambda}_n$ label the coefficients of the short-range region ($0 < R < R_\text{ref}$) and $C^{\lambda}_n$ the $R^{-n}$ terms in the long-range extrapolated domain ($R > R_\text{ref}$). For each angular dependency block, all coefficients and the $R_\text{ref}$ value were optimized within the fit. 

For each orientation, the fitted points resulted in good agreement with the corresponding \textit{ab initio} computed ones, with deviations on average within 1\% over the entire grid. 
The energy plots corresponding to the chosen \ce{H2} orientations are 
shown in Figure \ref{fig2}. 
It is apparent the weak anisotropy of the potential with respect to the $\{\theta^{\prime},\phi\}$ coordinates, thus validating the choice of truncating the potential to the $l_2\leq 2$ terms. 
For each orientation, the potential exhibits a minimum at $R$\,$\sim$\,$2.8$\,\AA\,and $\theta$\,$\sim$\,$112.5$\,degrees, i.e., with the \ce{H2} slightly leaning toward the phosphorous side of \ce{PO+} (see Figure~\ref{fig1}, where $R$ and $\theta$ have been purposely set to depict the minimum of the potential).  
 
\begin{figure}
\centering
 \includegraphics[scale=0.38]{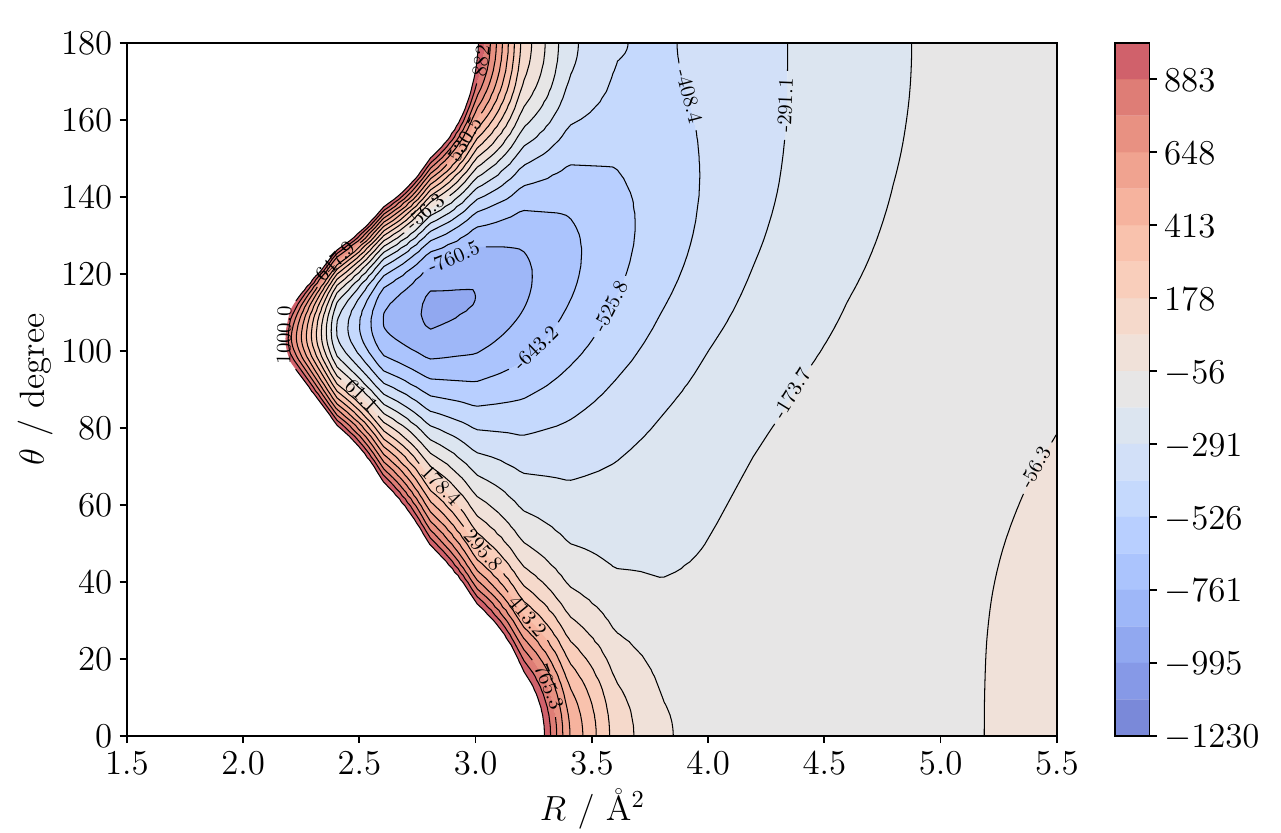}
 \caption{Contour plots of the averaged potential of \ce{PO+}--\ce{H2} over the five \{$\theta^{\prime},\phi$\} orientations (see Eq.~\eqref{vav}).}
 \label{fig3}
\end{figure}

The 4D potential of the system was finally retrieved by introducing a functional dependence on the four spherical harmonics (for the explicit dependence on the $\{\theta^{\prime},\phi\}$ coordinates the reader is referred to Equation 6 in \citet{tonolo2022hyperfine}):
\begin{equation}
\begin{aligned}
V &\left(R, \theta, \theta^{\prime}, \phi \right) = \, 2\, \sqrt{\pi} \,V_{\text{av}}(R,\theta)\, Y_{0 0}(\theta^{\prime}, \phi) +\\
+\,& 2\,\sqrt{\frac{\pi}{5}}\, \left[V(R, \theta, z)-V_{\text{av}}(R,\theta)\right] \,Y_{2 0}(\theta^{\prime}, \phi)+\\
+ & \, \sqrt{\frac{3 \pi}{10}}\, \left[V(R, \theta, a)-V(R, \theta, b)\right] \, [Y_{2 -1}(\theta^{\prime}, \phi) - Y_{2 1}(\theta^{\prime}, \phi)] + \\
+ & \, \sqrt{\frac{2 \pi}{15}}\, \left[V(R, \theta, x)-V(R, \theta, y)\right] \, [Y_{2 -2}(\theta^{\prime}, \phi) + Y_{2 2}(\theta^{\prime}, \phi)] \,.
\end{aligned}
\end{equation}
where $V_{\text{av}}(R, \theta)$ is the potential averaged over the five $\{\theta^{\prime},\phi\}$ orientations:
\begin{equation}\label{vav}
 \begin{aligned}
  V_{\text{av}}(R, \theta)&=  \frac{1}{7}[2\left(V(R,\theta,a)+V(R,\theta,b)\right)+ \\
  &+\left(V(R,\theta,x) +V(R,\theta,y)+V(R,\theta,z)\right)] \,,
 \end{aligned}
\end{equation}
for which a contour plot representation is shown in Figure \ref{fig3}.
The behavior of our potential is perfectly consistent with the trend of that computed by \citet{chahal2023po+} for the interaction between \ce{PO+} and \ce{He}, although --- as expected ---  the magnitude of the interaction is almost seven times higher in terms of energy: for the $\{\theta^{\prime}=90, \phi=0\}$ orientation with \ce{H2},
corresponding to the global (and unique) minimum of the potential, the energy is 1234.12\, \wn, while for the interaction with \ce{He} the minimum is located at 181.97\,\wn\@.

\subsection{Scattering Calculations}\label{scat}
\begin{figure}
 \begin{center}
  \includegraphics[scale=0.45]{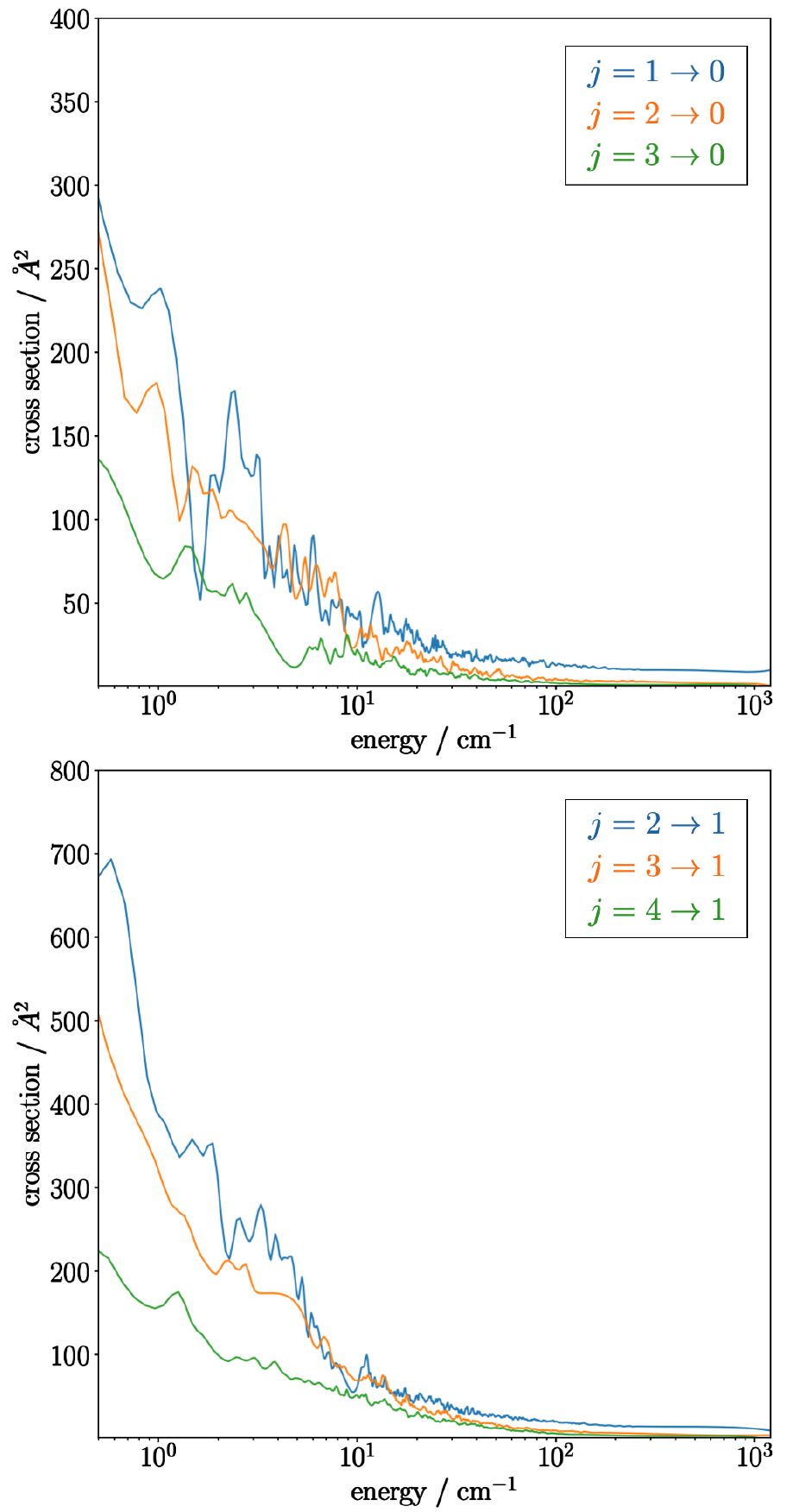}
 \end{center}
 \caption{Variation of some rotational de-excitation cross sections in the $2-1200$\,\wn energy range.}
 \label{fig7}
\end{figure}
\begin{figure}
 \begin{center}
  \includegraphics[scale=0.45]{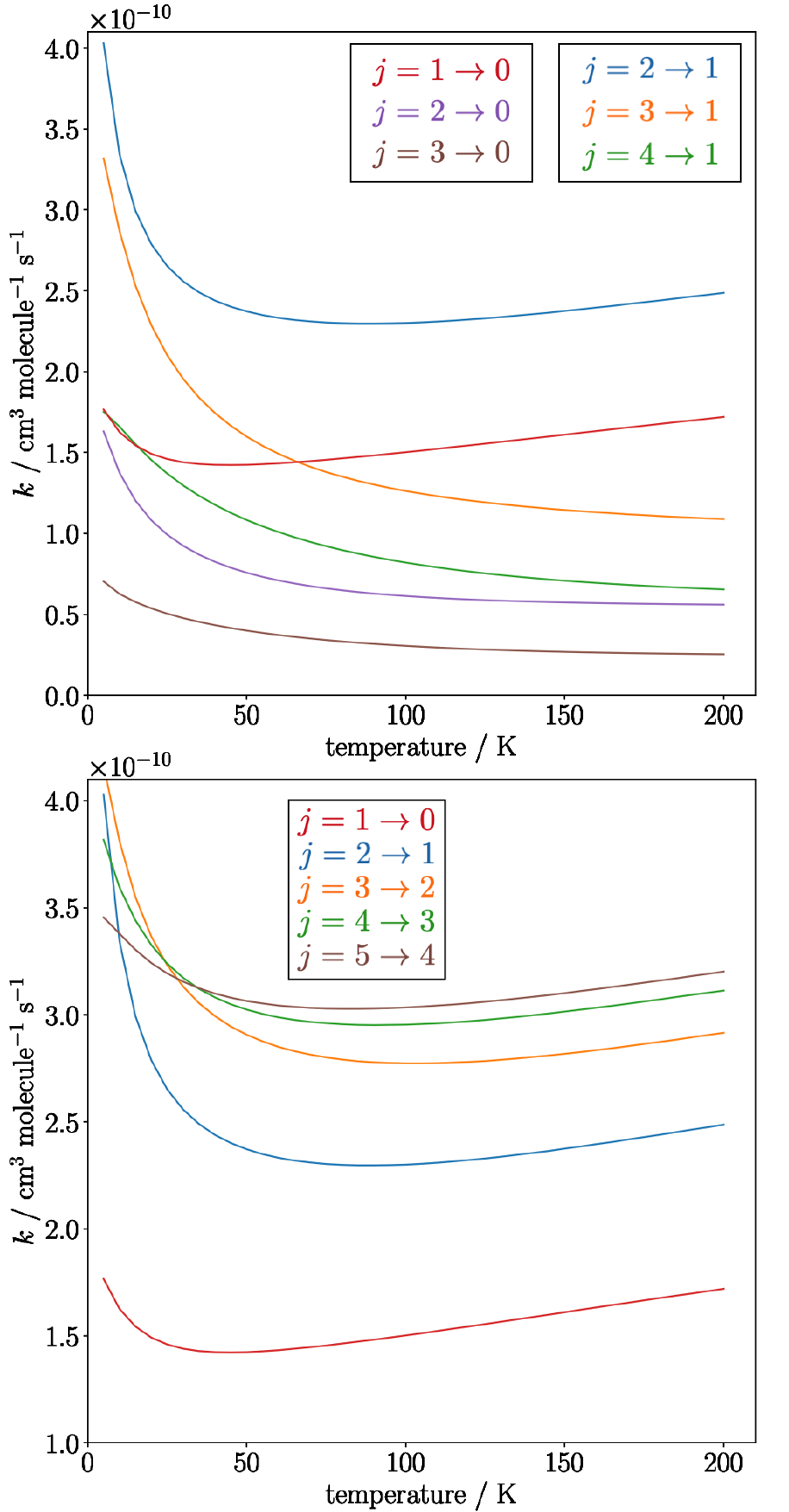}
 \end{center}
 \caption{Variation of some rotational de-excitation rate coefficients as a function of $T_{kin}$.}
 \label{fig8}
\end{figure}

\indent\indent
The second step in the derivation of the \ce{PO+}--\ce{H2} collisional rate coefficients is to perform dynamics calculations on top of the interaction potential. 
In this work, full quantum close-coupling (CC) calculations were carried out by employing the MOLSCAT program\footnote{\url{https://github.com/molscat/molscat}.} \citep{molscat14}.
Since our objective is to derive reliable state-to-state rate coefficients up to 200\,K, we thus targeted a total energy interval from 2\,\wn to 1200\,\wn.
Because of the irregular trends exhibited by the cross sections at low kinetic energies, we sampled with very fine steps (0.2\,\wn) the interval 2-- 50\,\wn, which have been then gradually increased up to 50\,\wn of step size above 500\,\wn. 

The angular part of the nuclear Schrödinger equation has been solved, and the radial dependence has then been modelled by numerical propagation. The employed propagator, also called LDMD/AIRY \citep{alexander1984hybrid,alexander1987stable,manolopoulos1986improved}, is an hybrid function that combines the Manolopoulos diabatic modified log-derivative (LDMD) propagator in the short range of $R$ with the Alexander-Manolopoulos Airy (AIRY) one for long $R$ distances. 
The former, indeed, uses a narrower propagation step when the energy gradient is higher, thus providing a good balance between stability and efficiency. The latter instead is particularly efficient at higher values of $R$, where a broader range needs to be covered, since it accounts for looser propagation steps.
We started the integration at $R=1.8$\,\AA\,and the switch point between the two propagators and the long range limit value were adjusted to ensure convergence within 2\% for the inelastic cross sections in the considered energy range.

The rotational basis adopted for \ce{PO+} included the first 32 rotational levels in the low energy range (2-50\,\wn), extended up to $j=50$ in the high-energy end of the scattering calculations (1200\,\wn).
The choice of the rotational basis of \ce{H2} requires some extra remarks.
The 4D potential defined above constitutes the best representation of the system as it also includes for the calculation of each cross section the influence of the $j>0$ rotational states of \ce{H2}. 
This, however, implies a major computational effort.

The impact of the coupling between $j=0$ and $j>0$ rotational states of \ce{H2} can be evaluated by performing a few comparative calculations of cross sections at $150$\,\wn using both the global 4D potential and the averaged 2D one, this latter being obtained from Eqs.~\eqref{pot} and~\eqref{vav} of the previous section.
The use of the 2D averaged potential, also known as the spherical approximation, is equivalent to treat the \ce{H2} projectile as a structureless species, i.e., behaving as a rotating sphere (\textit{para}-\ce{H2} with $j = 0$). Such approximation has already been found to be successful 
especially with ions and it significantly reduces the cost of the collision dynamics calculations \citep{spielfiedel2015new, balancca2020inelastic, cabrera2020relaxation, tonolo2022hyperfine}.
In the present case, the comparison between the 2D potential and the full 4D potential, in which the coupling with $j = 2$ rotational level was included, showed a mean average relative deviation of $\sim$\,$7.0$\% and a maximum discrepancy of 14.7\%. These deviations on the cross sections, reported in Table~\ref{tab1}, fall within the desired level of accuracy for astrophysical applications and justify the use of the spherical average potential approximation over the entire energy grid.
Given the physical conditions of the region where \ce{PO+} was observed, where --- despite the low densities --- the kinetic temperature is about 150\,K, collisions with \emph{ortho}-\ce{H2} should be considered too. 
For this reason, in Table~\ref{tab1}, a comparison of the cross sections obtained for collisions with \emph{para}-\ce{H2}($j=0$) and \emph{ortho}-\ce{H2}($j=1$) is also provided. Here, the mean average relative deviation is $\sim$\,$19$\%, with a maximum discrepancy of 48.24\%. Given such differences, we do not expect large discrepancies between the rate coefficients for collisions with \emph{ortho}- and \emph{para}-\ce{H2}, thus justifying the use of only \emph{para}-\ce{H2}($j=0$) in the rotational basis of our scattering calculations. This assumption is also in accordance with the results reported for other ions, e.g. \citet{walker2017inelastic, lara2019rotational,klos2011first, dagdigian2019interaction,desrousseaux2019collisional}.
\begin{table}
    \renewcommand{\tabcolsep}{0.8pt}
    \renewcommand{\arraystretch}{1.1}
        \scriptsize
        \caption{Computed cross sections 
        at $E = 150$\,\wn for \ce{PO+}$-$\,\ce{H2}$(j = 0)$ collisions obtained from the 2D spherically averaged potential and with the full 4D potential which includes the coupling with $j = 2$ rotational level of \ce{H2}. The comparison of the cross sections accounting for collisions between \emph{para}-\ce{H2} ($j=0$) and \emph{ortho}-\ce{H2} ($j=1$) is also reported. 
        }
        \footnotesize
        \centering 
        \begin{tabular}{lcdddcdd}
            \hline\hline
            \multirow{3}{*}{$j \rightarrow j^{\prime}$} & & \multicolumn{3}{c}{Cross sections / \AA$^2$} & & \multicolumn{2}{c}{\% Deviation}\\
            \cline{3-5} \cline{7-8}
            & &  \multicolumn{2}{c}{$j$=0} & \multirow{2}{*}{$\qquad j$=1} & & \multirow{2}{*}{$\qquad$ 2D/4D} & \multirow{2}{*}{\emph{o-/p-}} \\
            \cline{3-4}
            & &  \multicolumn{1}{c}{2D} & \multicolumn{1}{c}{4D}& & & & \\
            \hline
            1 $\rightarrow$ 0 &  & 33.10 & 30.46  & 33.84 & &  -8.68  \qquad &  2.19    \\                                    
            2 $\rightarrow$ 0 &  & 17.43 & 19.16  & 20.46 & &  9.02   \qquad &  14.81    \\                                    
            3 $\rightarrow$ 0 &  & 9.32  & 8.48   & 10.88 & &  -9.88  \qquad &  14.35    \\                                    
            4 $\rightarrow$ 0 &  & 12.64 & 11.89  & 14.39 & &  -6.29  \qquad &  12.17    \\                                    
            0 $\rightarrow$ 1 &  & 11.15 & 11.37  & 11.49 & &  1.90   \qquad &  2.93    \\                                                                                              
            2 $\rightarrow$ 1 &  & 26.61 & 24.55  & 29.11 & &  -8.42  \qquad &  8.57    \\                                    
            3 $\rightarrow$ 1 &  & 16.84 & 15.16  & 19.70 & &  -11.08 \qquad &  14.52    \\
            4 $\rightarrow$ 1 &  & 11.75 & 12.82  & 13.16 & &  8.38   \qquad &  10.72    \\
            0 $\rightarrow$ 2 &  & 3.60  & 3.30   & 4.27  & &  -8.99  \qquad &  15.75    \\                                                                                   
            1 $\rightarrow$ 2 &  & 16.31 & 15.11  & 17.21 & &  -7.94  \qquad &  5.22    \\                                            
            3 $\rightarrow$ 2 &  & 26.06 & 30.57  & 27.96 & &  14.73  \qquad &  6.79    \\                                            
            4 $\rightarrow$ 2 &  & 15.67 & 14.23  & 18.19 & &  -10.14 \qquad &  13.87    \\                                                                                                                               
            0 $\rightarrow$ 3 &  & 1.42  & 1.40   & 2.74  & &  -1.55  \qquad &  48.24    \\                                            
            1 $\rightarrow$ 3 &  & 7.62  & 8.14   & 11.87 & &  6.38   \qquad &  35.81    \\                                            
            2 $\rightarrow$ 3 &  & 19.24 & 18.81  & 22.66 & &  -2.30  \qquad &  15.10    \\ 
            4 $\rightarrow$ 3 &  & 27.30 & 28.84  & 31.06 & &  5.32   \qquad &  12.10    \\         
            0 $\rightarrow$ 4 &  & 1.57  & 1.70   & 2.88  & &  7.98   \qquad &  45.48    \\         
            1 $\rightarrow$ 4 &  & 4.33  & 4.14   & 7.61  & &  -4.44  \qquad &  43.13    \\
            2 $\rightarrow$ 4 &  & 9.42  & 9.69   & 13.58 & &  2.78   \qquad &  30.63    \\
            3 $\rightarrow$ 4 &  & 22.23 & 20.23  & 30.87 & &  -9.87  \qquad &  27.99    \\
            \hline
            \multicolumn{4}{l}{Average absolute \% deviation}  & &  & 7.30 & 19.02 \\
            \hline\hline
        \end{tabular}
        \label{tab1}
        \end{table}

The final calculations were thus performed using the 2D averaged potential, and the maximum value of the total angular momentum ($J$) was chosen to allow for convergence of the inelastic cross sections within 0.005\,\AA$^2$. 
The reduced mass ($\mu$) of the collisional system was set to 1.9327\,u, while the rotational energies of the two colliders were calculated from their equilibrium rotational and quartic centrifugal distortion constants. For \ce{PO+}, $B_e=0.787$\,\wn and $D_e=9.786\times10^{-7}$\,\wn were adopted \citep{petrmichl1991microwave}. 
For \ce{H2} molecule, the data given by \citet{huber1979abbreviated} were employed: $B_e=60.853$\,\wn and $D_e=4.71\times10^{-2}$\,\wn.

From the derived $\mathsf{S}$ matrix elements, the cross sections ($\sigma(E_{c})$) from an initial $j$ to a final $j^{\prime}$ state of \ce{PO+} at a given collision energy ($E_{c}$) can be retrieved. The energy dependence of some of them, involving the first rotational states of \ce{PO+}, as a function of $E_{c}$, is illustrated in Figure~\ref{fig7}.
Starting from inelastic ($j^\prime\neq j$) cross sections, the de-excitation rate coefficients, $k_{j \rightarrow j^\prime}(T)$, are straightforwardly derived by thermal averaging over the $E_c$:
\begin{equation}
\begin{aligned}
 k_{j \rightarrow j^{\prime}}(T) 
   &=\left(\frac{8}{\pi \mu k^{3} T^{3}}\right)^{1 / 2} \\
   &\times \int_{0}^{\infty} \sigma_{j \rightarrow j^{\prime}}\left(E_{c}\right) E_{c} \exp \left(-E_{c} / k T\right) \mathrm{d} E_{c} \,,
\end{aligned}
\end{equation}
where $k$ is the Boltzmann constant. 
We computed the (de)-excitation rate coefficients for the twenty lowest rotational levels of \ce{PO+} in the 5$-$200\,K temperature range. 
The complete set of them will be made available through the LAMDA \citep{schoier2010lamda, van2020leiden} and BASECOL \citep{dubernet2013basecol2012} databases. 
The plots in Figure~\ref{fig8} illustrate the temperature dependence of selected coefficients for which $j^{\prime}\rightarrow 0, j^{\prime}\rightarrow 1$ (upper plot) and $\Delta j=1$ (bottom plot).
A propensity toward transitions involving $\Delta j = 1$
clearly stands out, while it decreases with the increment of $\Delta j$.
In addition, inelastic rate coefficients exhibit a propensity toward transitions involving $j$ levels with the highest
multiplicity although, at low temperatures ($T<50\,K$), this preference may reverse. 
This pattern is in accordance with the one observed for the isoelectronic species \ce{NO+} with \emph{para}-\ce{H2} \citep{cabrera2020relaxation}. This agreement is also reflected in the comparison of the values of the rate coefficients for the two species. 
The recent collisional study on the \ce{PO+/He} system \citep{chahal2023po+}, moreover, provides us a basis to test the suitability of using \ce{He} to simulate the behavior of \emph{para}-\ce{H2}. The mass scaled rate coefficients \citet{schoier2005atomic} for the collision of \ce{PO+} with \ce{He}, however, resulted to highly underestimate the values obtained with \ce{H2}, leading to discrepancies up to one order of magnitude in several cases. In addition, the two sets of rate coefficients exhibit different behaviors at temperatures below 50\,K, including also divergent propensity rules. 
Since these discrepancies are particularly evident under the typical ISM conditions, only collisional data with \ce{H2} provide the most reliable means for astrophysical applications. 
Indeed, it is well established \citep{yazidi2014revised,denis2019rotational,bop2019hyperfine,cabrera2020relaxation} that the inaccuracy of \ce{He} as a template of the \ce{H2} perturber is particularly pronounced in the case of ions because of the different behavior of the long range part of the potential for the \ce{ion-He} and \ce{ion-H2} interactions.

\section{Astrophysical Applications}
\label{aa}
\begin{figure}  
 \includegraphics[scale=0.28]{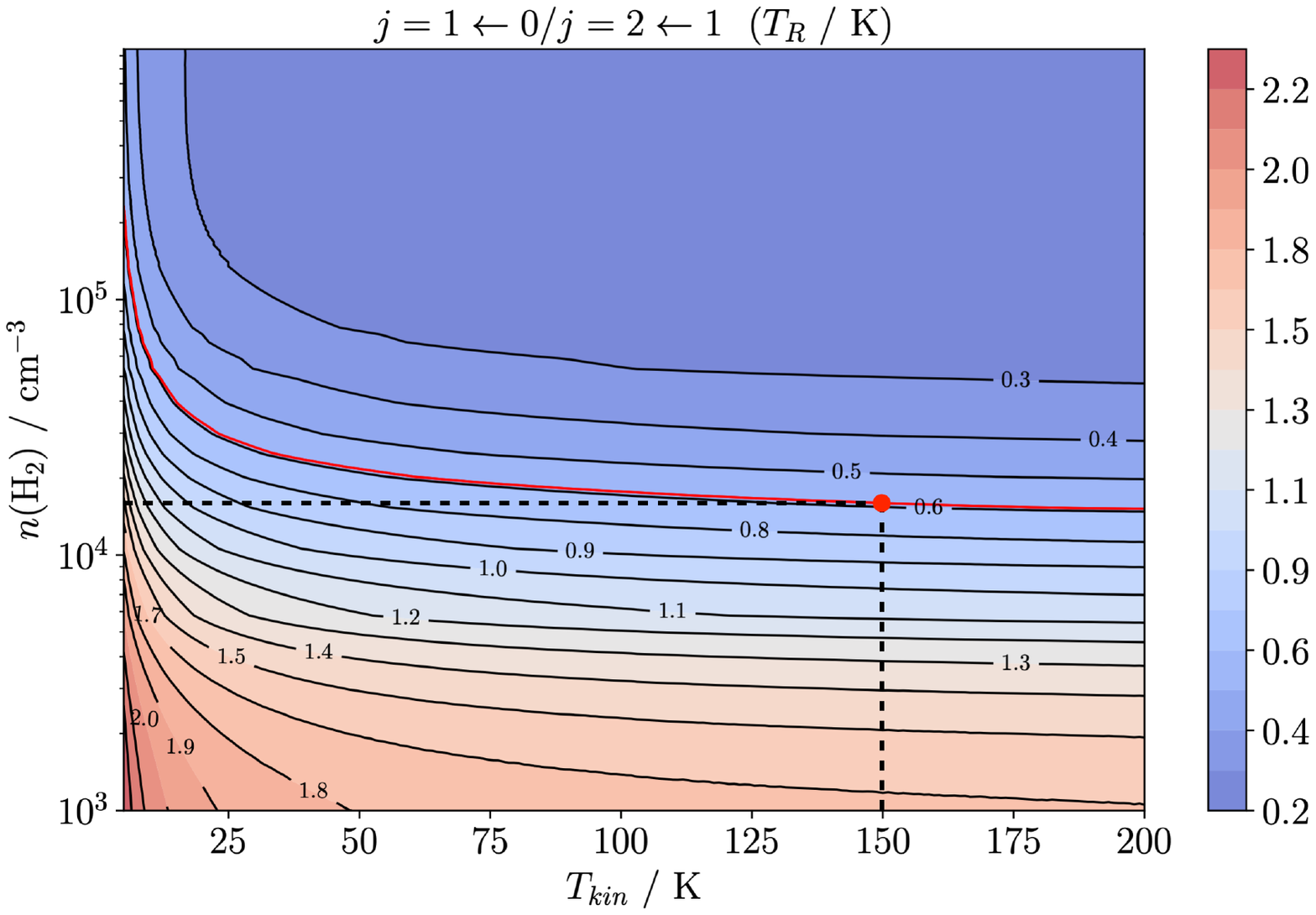}
 \caption{Contour plot showing the variation of the intensities ratio $j=1-0/j=2-1$ with respect to the change of $T_\text{kin}$ ($x$ axis) and $n({\ce{H2}})$ ($y$ axis). The isocurve of the intensity ratio observed by \citet{rivilla2022ionize} is highlighted in red.}
 \label{fig4} 
\end{figure}
\begin{figure*}  
 \includegraphics[scale=0.60]{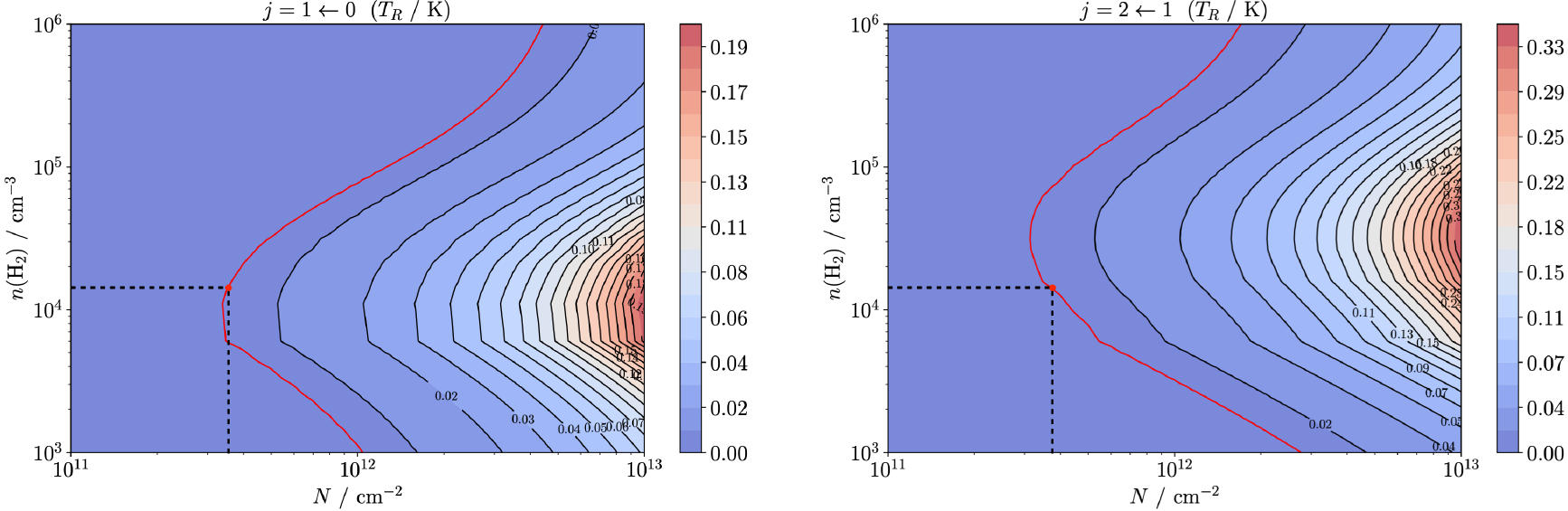}
 \caption{Contour plot showing the variation of the $j=1-0$ and $j=2-1$ intensities with respect to the change of the column density ($x$ axis) and $n({\ce{H2}})$ ($y$ axis). The isocurves of the two intensities observed by \citet{rivilla2022ionize} are highlighted in red.}
 \label{fig5} 
\end{figure*}

\indent\indent
Our new collisional coefficients provide a good basis to test the suitability of the LTE approximation to model the transitions detected by \cite{rivilla2022ionize} in the G+0.693-0.027 molecular cloud. Due to the relatively low density of this source (several $1\times10^{4}$\,cm$^{-3}$; \citet{zeng2020cloud}), the energy levels of the molecules are not thermalized at the kinetic temperature of the cloud ($\sim$\,150\,K; \citet{zeng2018complex}), but they reach a ``quasi-thermalization'' at an excitation temperature ($T_{\text{ex}}$) that is significantly lower than $T_{\text{kin}}$ (see detailed explaination of this effect in \citet{goldsmith1999population}). In absence of collisional data, LTE approach is hence used, giving typical $T_{\text{ex}}$ in the range of $5-20$\,K (see \citet{zeng2018complex}).
The spectral survey in \citet{rivilla2022ionize} covered four different rotational transitions of \ce{PO+}, though, only the $j=1-0$ and $j=2-1$ ones appear free from contamination by other species.
This implies the system to have only two degrees of freedom for the rotational population modelling, thus the retrieved physical parameters are to be viewed with some caution.

We performed radiative transfer calculations using the RADEX code\footnote{\url{https://home.strw.leidenuniv.nl/~moldata/radex.html}.} \citep{van2007computer}. 
Since the structure and dynamics of G+0.693-0.027 cloud is not well constrained, the source geometry was approximated to a static sphere of uniform density.
We also assumed $T_\text{CMB}=2.73\,$K as background temperature, while the line width was set at $18\,$km/s, in accordance with observations \citep{rivilla2022ionize}.
State-to-state collisional coefficients involving the first twenty rotational levels of \ce{PO+}, in the temperature range from 5\,K to 200\,K, and a \ce{PO+} dipole moment of 3.13\,debye \citep{rivilla2022ionize} have been employed for the calculation of the Einstein $A$ coefficients of the corresponding $(j+1)-j$ radiative transitions.

At first, we performed a preliminary test of the variation of the ($j=1-0$)/($j=2-1$) intensity ratio 
as a function of both the kinetic temperature ($T_\text{kin}$) and the density of hydrogen ($n({\ce{H2}})$). The result is illustrated in Figure~\ref{fig4}, where the isocontours corresponding to the observed intensity ratio (0.64) have been marked in red. 
It is obvious that the constraint of the $T_\text{kin}$ on $n({\ce{H2}})$ is very weak, particularly when $T_\text{kin}>100$\,K. Such a mild dependence allows us to retrieve a reliable estimate of the gas density of G+0.693-0.027, irrespective of the possible inaccuracies of $T_\text{kin}\sim$\,$150$\,K.
In fact, from the observed intensity ratio we obtain $n(\ce{H2})\sim$\,$1.5\times10^4$\,cm$^{-3}$, in good agreement with the value previously estimated by \citet{zeng2020cloud} of $\sim$\,$1\times10^4$\,cm$^{-3}$.

Going into the analysis of the single transitions, we plot in Figure~\ref{fig5} the line intensity trends at 150\,K as a function of $n(\ce{H2})$ and the column density ($N$). Again, the isocontours corresponding to the observed results,
i.e. 7\,mK for $j=1-0$ and 11\,mK for $j=2-1$,\ are highlighted in red.
Both plots substantially agree in indicating that, for $n(\ce{H2})\sim$\,$1.5\times10^4$\,cm$^{-3}$, the observed line intensities imply a \ce{PO+} column density of $\sim$\,$3.7\times10^{11}$\,cm$^{-2}$, a value which is $\sim$\,$60\%$ (a factor of $\sim$\,$1.7$) lower than that retrieved from LTE assumption \citep{rivilla2022ionize}. The reason of this slight difference is that the LTE model tends to underestimate
the population of the higher-energy levels, therefore requiring a higher column density to reproduce the observed intensities.
The decrease in the actual value of the column density of \ce{PO+} is also reflected in a diminution of the $N(\ce{PO+})/N(\ce{PO})$ ratio derived by \citet{rivilla2022ionize}, which becomes $\sim$\,$0.072$. This leads to a reduction in the \ce{PO} ionization rate previously predicted, which nevertheless remains predominant with respect to those retrieved for \ce{NO} and \ce{SO}.
\begin{figure*}  
 \includegraphics[scale=0.60]{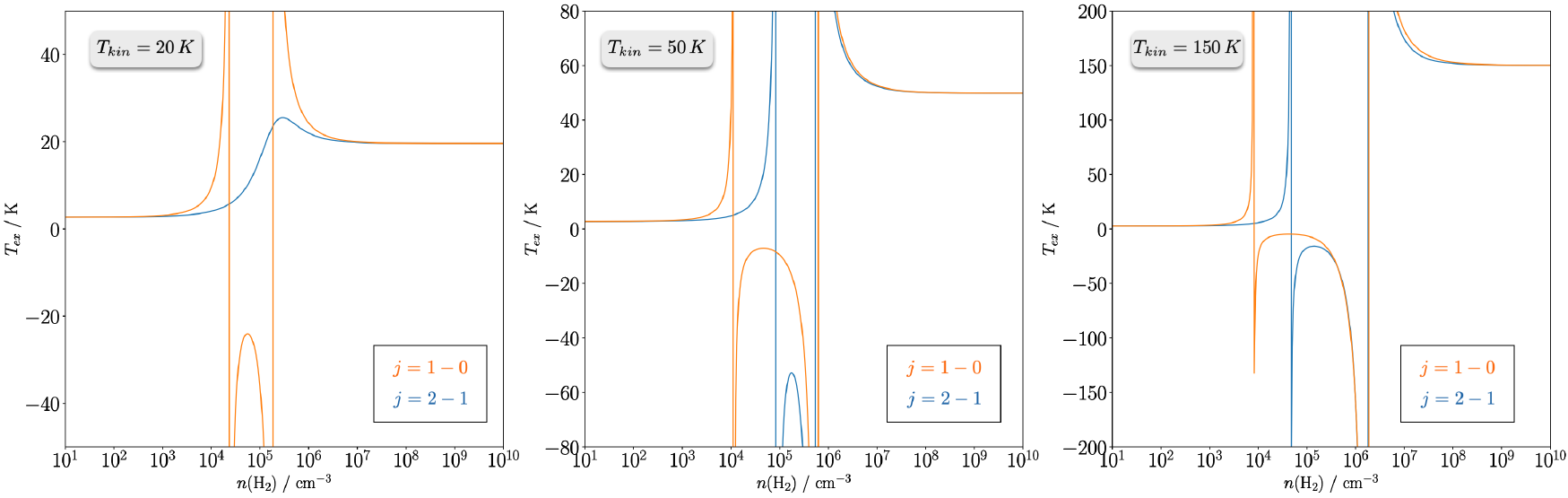}
 \caption{Variation of excitation temperature as a function of $n(\ce{H2})$ for the first two transitions of \ce{PO+} at $T_\text{kin}=$\,20,\,50 and 150\,K.}
 \label{fig6} 
\end{figure*}
In any case, the outcome of our analysis is that the LTE approximation to model low-$j$ lines of \ce{PO+} provides reasonable results when applied to the physical conditions of G+0.693-0.027. As has already been found (e.g., see discussion in \citet{colzi2022deuterium}), indeed, in G+0.693-0.027 the ``quasi-thermalization'' condition is often fulfilled, with the distribution of the population among the levels being adequately described by only one $T_\text{ex}$. This results in a discrete consistency between the $N$(non-LTE) and the $N$(LTE). 
Nevertheless, an important advantage of non-LTE analysis is that it allows to constrain the gas density, thus providing a means to validate the previous modelling predictions.

Still, in view of future observation of the \ce{PO+} ion in different sources, it is interesting to explore a wider range of densities and kinetic temperatures.
To this aim, we present in Figure~\ref{fig6} the trend of the $T_\text{ex}$ of the 
$j = 1 - 0$ and $j = 2 - 1$ lines as function of $n(\ce{H2})$ for three different values of $T_\text{kin}$.
It can be seen that, for densities around $10^5$\,cm$^{-3}$, maser phenomena are predicted, because of an inversion of the population between the levels. 



\section{Conclusions}
\label{concl}
\indent\indent
The present study was triggered by the recent detection of \ce{PO+} in the ISM \citep{rivilla2022ionize}; its aim is to support the interpretation of the present and future observations of this ion by providing an accurate description of its collision with the main astrochemical perturber \textit{para}-\ce{H2}.
We characterized the interaction PES by computing over 3250 \emph{ab initio} points using the CCSD(T)-F12a/aug-cc-pV(Q+d)Z level of theory in the four $\{R,\theta,\theta^{\prime},\phi\}$ Jacobi coordinates. 
Subsequently, we fitted the potential as an expansion over angular functions. Before performing scattering calculations, we assessed the coupling effects between the $j=0,2$ rotational states of \ce{H2}, which showed a minor impact on the cross sections. This allowed us to significantly simplify the scattering calculations by employing a spherical approximation of the potential averaged over five different orientations of \ce{H2}. Also, the comparison of some values of the cross sections between \textit{para}-\ce{H2} and \textit{ortho}-\ce{H2} revealed a good collisional agreement of the two species. This, as exhibited for other ions too \citep{walker2017inelastic,lara2019rotational,klos2011first,dagdigian2019interaction,desrousseaux2019collisional}, suggests that the obtained results with \emph{para}-\ce{H2} are adequate to describe the collisional behavior of \ce{PO+} even at high temperatures, where the influence of \emph{ortho}-\ce{H2} may also have an impact. 
The state-to-state collisional coefficients between the twenty lowest rotational levels of \ce{PO+} and for temperatures ranging from 5 to 200\,K were thus derived. 
A comparison with the data recently obtained by \citealt{chahal2023po+} for the \ce{PO+} and \ce{He} collisional system revealed large discrepancies in the rate coefficients values, up to one order of magnitude in several cases, even despite the mass scaled contribution. This proves, for astrophysical purposes, a scarce reliability of \ce{He} to simulate the behavior of \ce{H2} as colliding perturber of \ce{PO+}.

Finally, the computed collisional dataset allowed us to refine the abundance of \ce{PO+} measured in the G+0.693-0.027 cloud, which resulted quite consistent with the one derived with LTE approximation (only a factor of $\sim$\,1.7 lower). Moreover, the derived \ce{H2} density of the gas ($1.5\times10^4$\,cm$^{-3}$) resulted in good agreement with previous estimates \citep{zeng2020cloud}.
Radiative transfer calculations revealed maser behavior for the first rotational transitions of \ce{PO+} at different $T_\text{kin}$ for densities around $10^4-10^6$\,cm$^{-3}$. 
These conditions encompass a significant portion of the interstellar sources, 
thus foregrounding the importance of the computed collisional coefficients to ensure an accurate modelling of the abundance of \ce{PO+} in the ISM.

\section*{Acknowledgements}
This work has been supported by MUR (PRIN Grant Number 202082CE3T) and by the University of Bologna (RFO funds). 
The COST Action CA21101 ``COSY - Confined molecular systems: from a new generation of materials to the stars’’ is also acknowledged.
Moreover, we acknowledge financial support from the European Research Council (Consolidator Grant COLLEXISM, Grant Agreement No. 811363).
François Lique acknowledges financial support from the Institut Universitaire de France and the Programme National “Physique et Chimie du Milieu Interstellaire” (PCMI) of CNRS/INSU with INC/INP cofunded by CEA and CNES.
V.M.R. acknowledges support from the project RYC2020-029387-I funded by MCIN/AEI/10.13039/501100011033.

\section*{Data Availability}

The data underlying this article will be made available through the LAMDA \citep{schoier2010lamda, van2020leiden} and BASECOL \citep{dubernet2013basecol2012} databases.
 


\bibliographystyle{mnras}
\bibliography{PO+} 




%
%


\bsp	
\label{lastpage}
\end{document}